\documentclass[intlimits,sumlimits,12pt]{iopart}

\expandafter\let\csname equation*\endcsname\relax
\expandafter\let\csname endequation*\endcsname\relax

\usepackage{iopams}
\usepackage{amsmath}
\usepackage{dsfont}
\usepackage{color}
\usepackage{comment}
\usepackage[numbers]{natbib}

\usepackage{hyperref}

\usepackage{graphicx}
\usepackage{amsfonts}
\usepackage{tikz}
\usepackage{framed,color}
\definecolor{shadecolor}{rgb}{1,0.8,0.3}
\usepackage{setspace}
\usepackage{dashrule}
\usepackage{longtable}
\usepackage{slashed}

\newcommand{\Pf}{{\text{Pf}}}

\newcommand{\eins}{\leavevmode\hbox{\small1\kern-3.8pt\normalsize1}}

\begin{document}

\title[]{Statistical Topology --- Distribution and Density Correlations
  of Winding Numbers in Chiral Systems}

\vspace{1cm}

\hspace{-0.85cm}\textit{\large Dedicated to Giulio Casati on the Occasion of His 80th Birthday.}

\vspace{0.2cm}

\author{Thomas Guhr}

\address{Fakult\"at f\"ur Physik, Universit\"at Duisburg--Essen, Duisburg, Germany}

\ead{thomas.guhr@uni-due.de}


\begin{abstract}
Statistical Topology emerged since topological aspects continue to
gain importance in many areas of physics. It is most desirable to
study topological invariants and their statistics in schematic models
that facilitate the identification of universalities. Here, the
statistics of winding numbers and of winding number densities are
addressed.  An introduction is given for readers with little
background knowledge. Results that my collaborators and I obtained in
two recent works on proper random matrix models are reviewed, avoiding
a technically detailed discussion. A special focus is on the mapping
of topological problems to spectral ones as well as on the first
glimpse on universality.
\end{abstract}

\vspace{2pc}





\section{Introductory Remarks}
\label{sec1}

Statistical Topology aims at combining, in a generalizing form,
topological questions appearing in physics with the powerful concepts
of Random Matrix Theory (RMT) which is capable of describing spectral
statistics in a huge number of systems, stemming from different areas
of physics and beyond. The focus in this work is exlusively on winding
numbers and associated statistical quantities studied in the framework
of a random matrix model, other topological invariants which are also
of considerable interest are not discussed. The long--term aim is to
study the emergence of universalities whose identification and usage
is always, in all branches of statistical physics, the most rewarding
enterprise.  I have two goals: First, I want to present an
introduction to Statistical Topology, restricted to statistical
problems which are related to winding numbers, for readers without
pertinent background. Neither physics expert jargon nor heavy
mathematics and mathematical physics terminology are used. Second, I
want to review and summarize results that my collaborators and I
obtained in two recent studies~\cite{Braun2022,Hahn2022}. We calcuated
for a chiral unitary random matrix model correlators of winding number
densities and the winding number distribution. We also computed
generators for these correlatores in a chiral unitary and a chiral
symplectic random matrix model. Furthermore, we made first steps to
find universalities.

The paper is organized as follows: in Section~\ref{sec2} the salient
features of winding numbers and chiral symmetry are presented. In
Section~\ref{sec3} a schematic model with the necessary mathematical
setup is formulated. Results are reviewed in Section~\ref{sec4},
discussion and conclusions are given in Section~\ref{sec5}.

\section{Winding Number and Chirality}
\label{sec2}

After briefly revisiting the occurence of winding numbers in complex
analysis in Section~\ref{sec21}, the Kitaev chain is discussed in
Section~\ref{sec22} and the statistics ansatz is motivated in
Section~\ref{sec23}. The research is put in the framework of Quantum
Chromodynamics (QCD) and Condensed Matter Physics in
Section~\ref{sec24}, summarizing the corresponding remarks in
Ref.~\cite{Braun2022,Hahn2022}.

\subsection{A Simple Topological Invariant in Complex Analysis}
\label{sec21}

\begin{figure}[htbp]
  \begin{center}
  \includegraphics[width=0.4\textwidth]{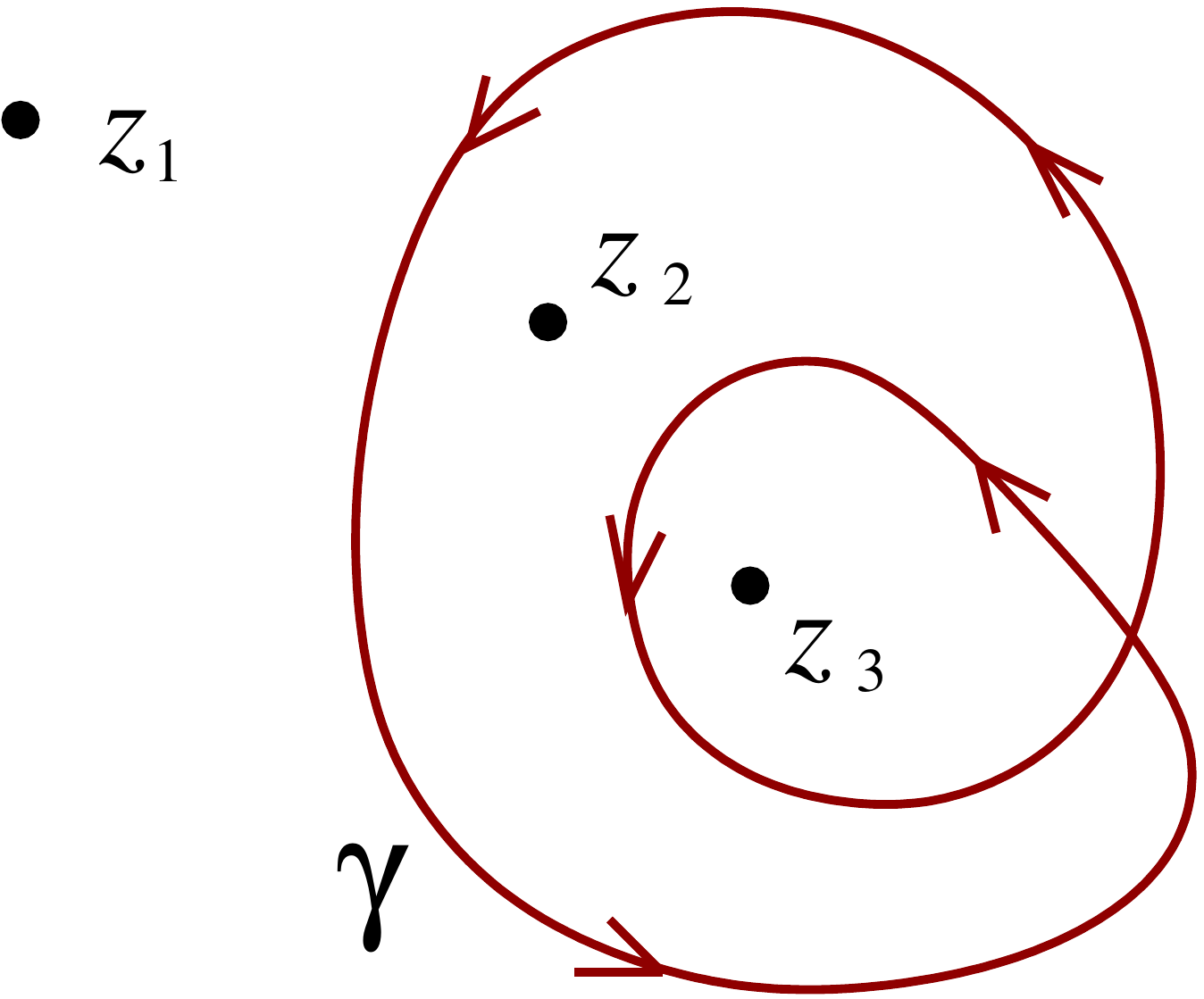}\hfill
  \includegraphics[width=0.4\textwidth]{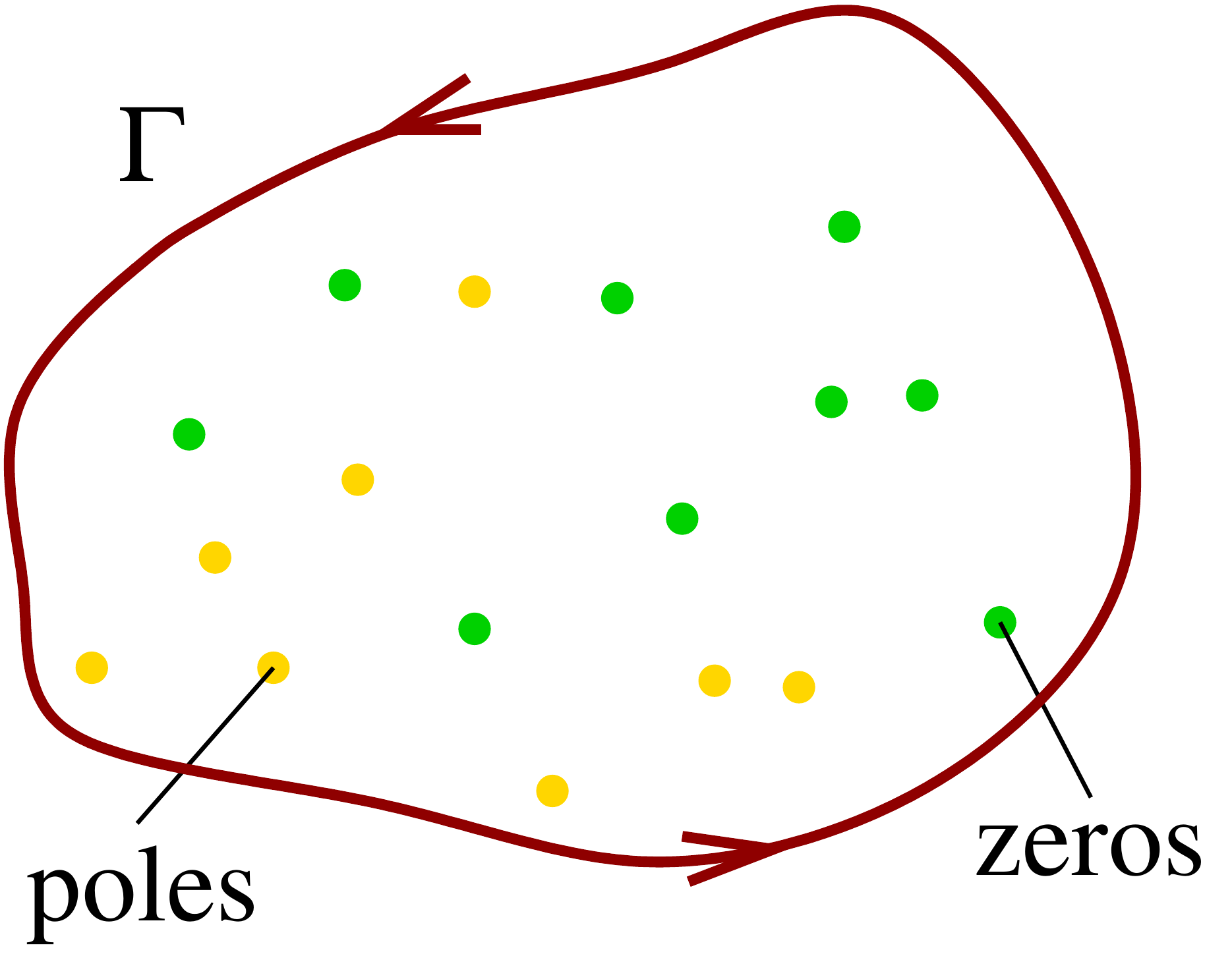}
  \end{center}
  \caption{Left: Three points $z_i, \ i=1,2,3$ in the complex plane
    $\mathbb{C}$ and a closed contour $\gamma$.  Right: A closed
    contour $\Gamma$ encircling zeros and poles of a meromorphic
    function $f(z)$.}
  \label{fig1}
\end{figure}

The winding number is a topological concept encountered in complex
analysis. Before discussing applications in physics, we briefly sketch
the mathematical background.  The winding number $W=W(z_i)$ counts,
how many times a point $z_i$ in the complex plane $\mathbb{C}$ is
encircled by a closed contour $\gamma$, where counterclockwise or clockwise give
a positive or a negative contribution, respectively. An example is
shown in Figure~\ref{fig1}, we have $W(z_1)=0$, $W(z_2)=1$ and
$W(z_3)=2$. Obviously, the winding number $W(z_i)$ is a topological
constant or, in physics terminology, a quantum number. It is invariant
under all deformations of $\gamma$ that do not cross the point $z_i$
in question. In particular, the winding number is always a positive or
negative integer, $W\in \mathbb{Z}$. It may be written as the contour
integral
\begin{equation}
  W(z_i) = \frac{1}{2\pi i} \oint\limits_\gamma \frac{d\zeta}{\zeta-z_i} \ .
\label{eq1.1} 
\end{equation}
One easily establishes the link to Cauchy's argument principle:
Consider a meromorphic function $f(z)$ and a closed contour $\Gamma$,
encircling some zeros and poles of $f(z)$ in the complex plane
$\mathbb{C}$ as shown in the example in Figure~\ref{fig1}. The
integral along this contour $\Gamma$ over the logarithmic derivative
of $f(z)$ yields the difference of the number $N_Z$ of zeros and the
number $N_P$ of poles, hence
\begin{equation}
\frac{1}{2\pi i} \oint\limits_\Gamma \frac{f'(z)}{f(z)} dz = N_Z - N_P \ .
\label{eq1.2} 
\end{equation}
The close relation to the winding number is found by making the change
of variable $\zeta=f(z)$ and accordingly of the contour,
$\Gamma\rightarrow\ f(\Gamma)$,
\begin{equation}
N_Z - N_P = \frac{1}{2\pi i} \oint\limits_\Gamma \frac{f'(z)}{f(z)} dz
  = \frac{1}{2\pi i} \oint\limits_{f(\Gamma)} \frac{d\zeta}{\zeta}
  = W(0) \ .
\label{eq1.3} 
\end{equation}
We conclude that $N_Z-N_P$ is the winding number $W(0)$ of the closed
contour $f(\Gamma)$ around the origin $z=0$. As, from now on, all
winding numbers will refer to the origin, we drop the argument and
simply write $W$.

\subsection{Kitaev Chain and Winding Number}
\label{sec22}

To illustrate the occurence of topological invariants in physics, we
\begin{figure}[htbp]
  \begin{center}
  \vspace{0.4cm}
  \includegraphics[width=0.6\textwidth]{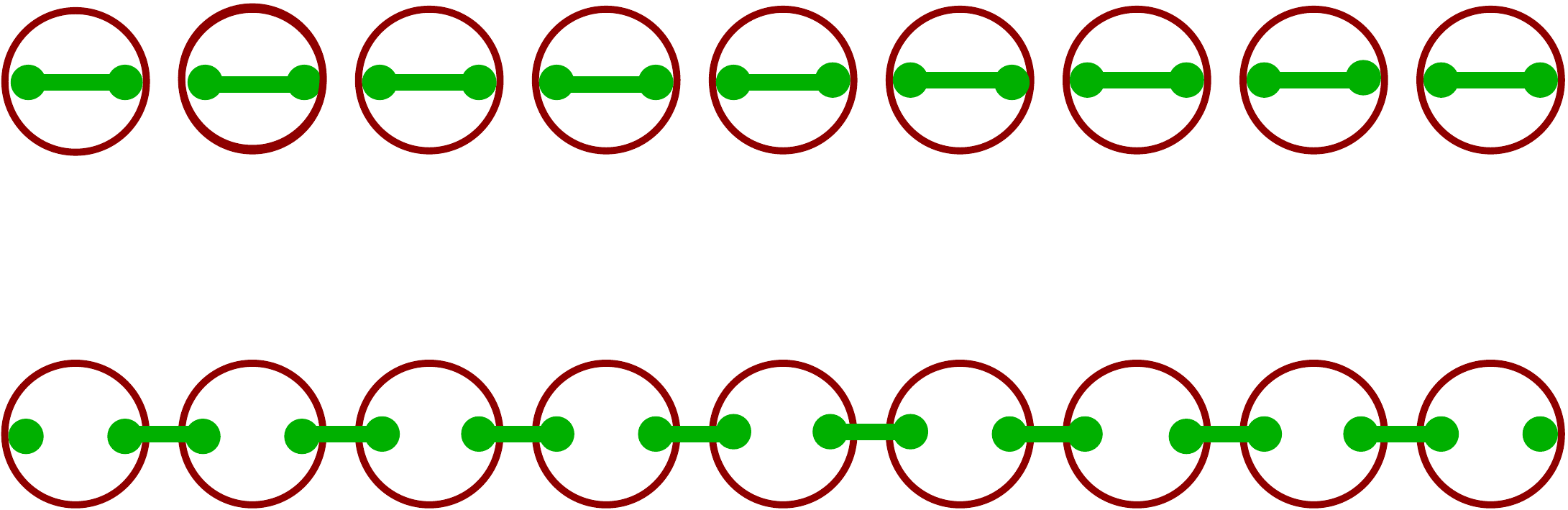}
  \vspace{0.4cm}
  \end{center}
  \caption{Kitaev chain, electrons as larger open circles (red),
    Majorana fermions as small dots (green) with the pairing indicated
    by connecting lines (green). Top: all Majorana fermions are
    paired, normal or trivial superconducting phase.  Bottom: unpaired
    Majorana fermions at the ends of the chain, topological
    superconducting phase.}
  \label{fig2}
\end{figure} 
look at the Kitaev chain~\cite{Kitaev2000,Kitaev2001} as a
prominent example.  It consists of spinless electrons with
next--neighbour hopping and superconductive pairing. The Hamiltonian
reads, in a slightly simplified form sufficient for the present
discussion,
\begin{equation}
  \hat{H} = \displaystyle\sum\limits_{n} \biggl(t\left(\hat{c}^\dagger_n\hat{c}_{n+1}
             +\hat{c}^\dagger_{n+1}\hat{c}_n\right) + \mu \hat{c}^\dagger_n\hat{c}_n 
    +      \frac{\Delta}{2} \left(\hat{c}^\dagger_{n+1}\hat{c}^\dagger_{n} + \hat{c}_{n}\hat{c}_{n+1}\right) \biggr) \ ,
\label{eq1.4} 
\end{equation}
where $\hat{c}_n$ and $\hat{c}^\dagger_n$ are annihilation and
creation operators, respectively, at position $n$ on the chain.
Moreover, $\mu$ and $\Delta$ are chemical and pairing potentials and
$t$ the hopping strength. The Hamiltonian may be reformulated in terms
of Majorana fermions whose number is twice that of the
electrons. Remarkably, depending on the parameters, there are two
possibilities, as schematically depicted in Figure~\ref{fig2}.  Either
all Majorana fermions are paired or, at the ends of the chain, two of
them are unpaired \cite{PPD2022}. In the former case the chain is in a
normal or trivial superconducting phase, in the latter in a
topological one.  This aspect deserves further discussion.

In Fourier space, the Kitaev chain corresponds to the
Bloch--Bogolyubov--de Gennes Hamiltonian matrix $H(k)$. It is a
crucial that this $2\times 2$ matrix satisfies chiral symmetry,
\begin{eqnarray}
  \{H(k),\mathcal{C}\} = 0 \qquad \text{with} \qquad
\mathcal{C}=\begin{bmatrix}
1 & 0
\\
0 & -1
\end{bmatrix} \ .
\label{eq1.6} 
\end{eqnarray}
The matrix $\mathcal{C}$ is the chiral operator in its proper basis
and $\{\ ,\ \}$ is the anticommutator. It is then possible to
write the  Hamiltonian matrix in the form
\begin{eqnarray}
H(k) = \vec{d}(k)\cdot\vec{\sigma} \qquad \text{with} \qquad \vec{d}(k)=(0, \Delta\sin k, \mu + 2t \cos k) \ .
\label{eq1.5} 
\end{eqnarray}
Hence, using the three--component vector $\vec{\sigma}$ of the
$2\times 2$ Pauli matrices, $H(k)$ is found to be a scalar product
with all
\begin{figure}[htbp]
  \begin{center}
  \vspace{0.4cm}
  \includegraphics[width=0.8\textwidth]{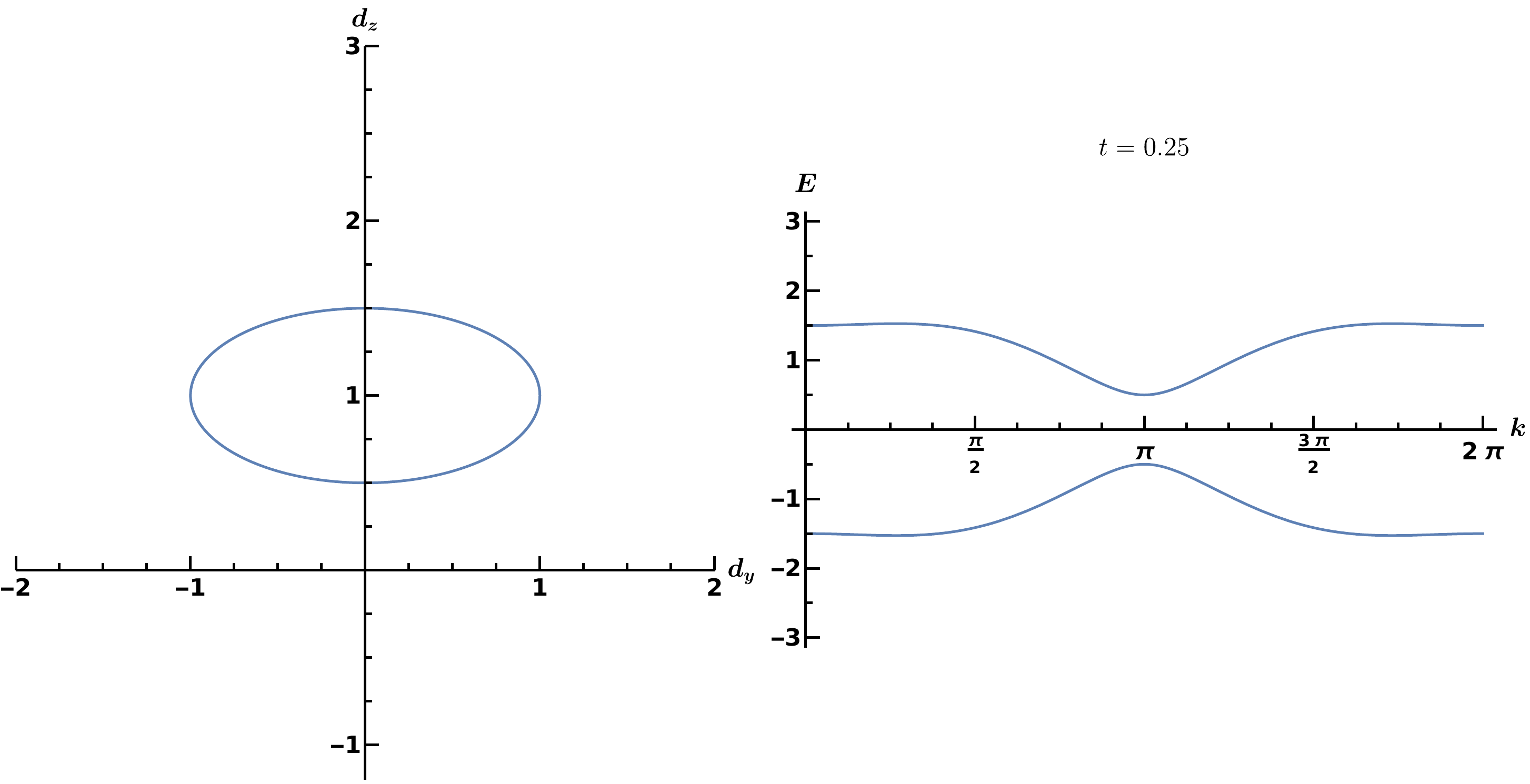}\\
  \vspace{0.2cm}
  \includegraphics[width=0.8\textwidth]{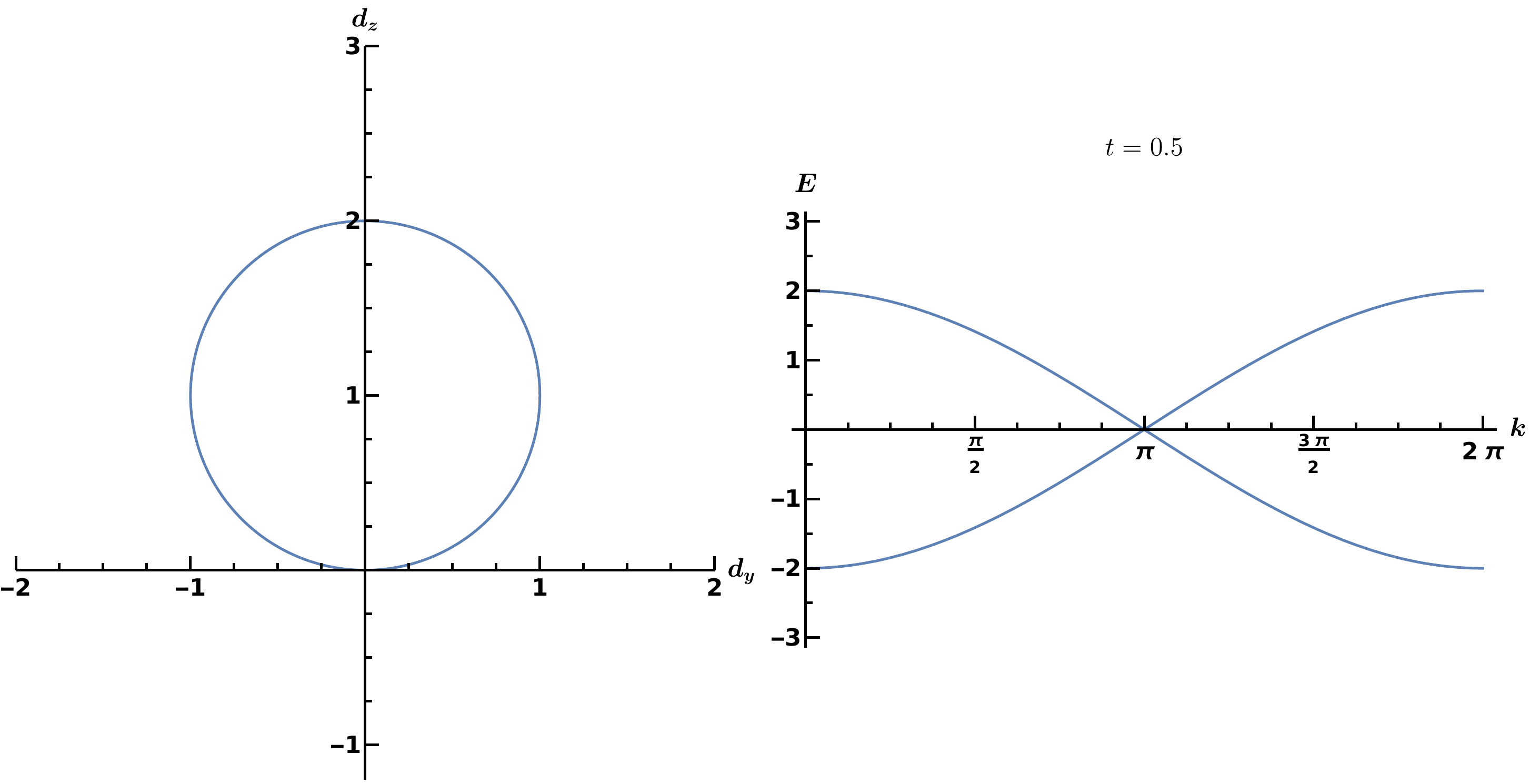}\\
  \vspace{0.2cm}
  \includegraphics[width=0.8\textwidth]{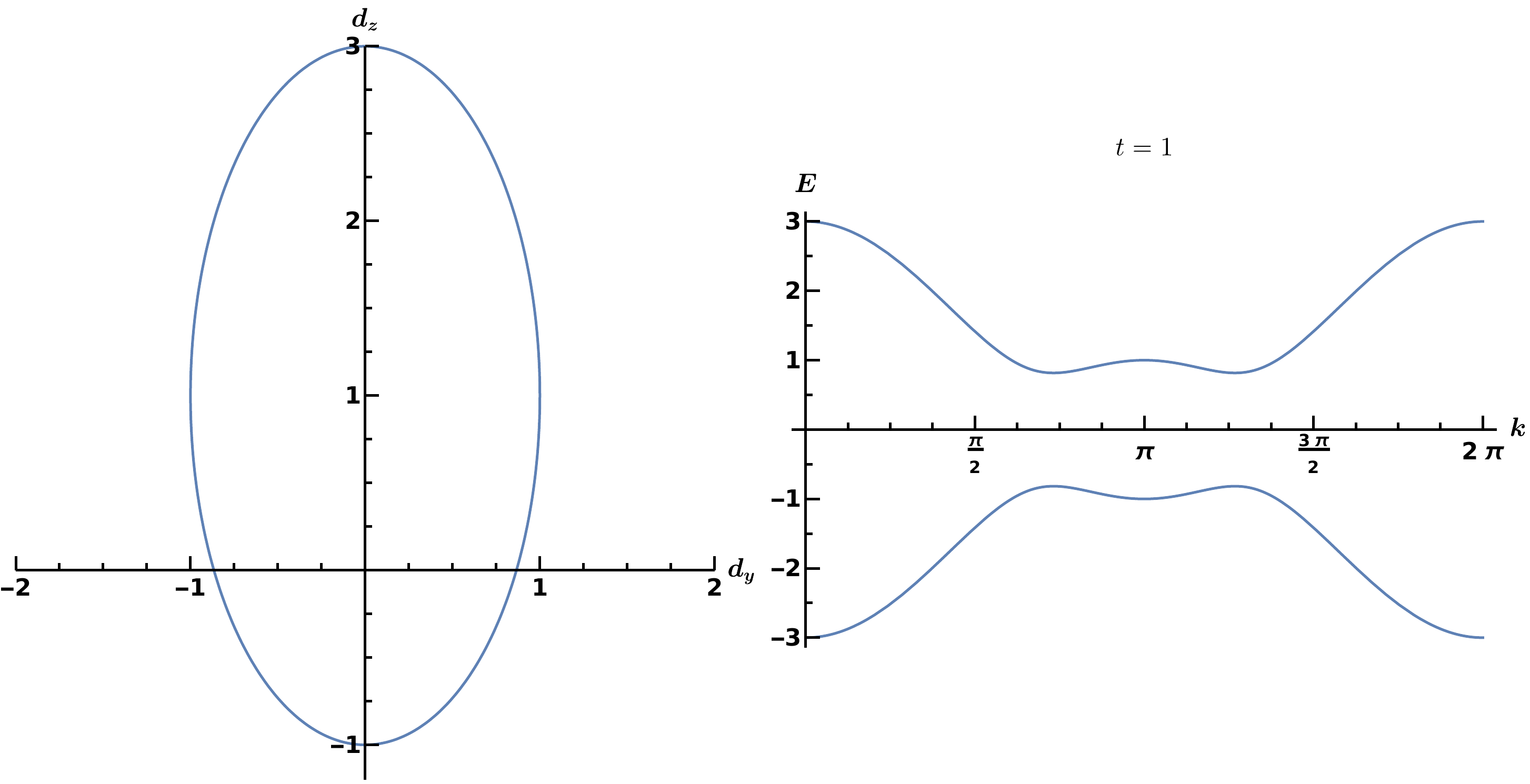}
  \vspace{0.4cm}
  \end{center}
  \caption{Ellipses described by $\vec{d}(k)$ (left) and corresponding
    dispersion relations $E(k)$ (right).  Top: $t=0.25$, normal
    superconducting phase, $W=0$.  Center: $t=0.5$, phase transition
    point.  Bottom: $t=1$, topological superconducting phase,
    $W=1$. Courtesy of Nico Hahn.}
  \label{fig3}
\end{figure} 
physics encoded in the vector $\vec{d}(k)$ that depends on the wave
number $k$ and the three parameters $\mu$, $\Delta$ and
$t$. Importantly, the first component is zero, $d_x=0$.  This
restriction to effectively only two dimensions can be shown to be a
consequence of chiral symmetry \eqref{eq1.6}.

To see how topology enters, we notice that the vector $\vec{d}(k)$
describes an ellipse with parameter $k$ on the curve, $\mu$ determines
the postion of its center, $\Delta$ and $t$ determine its shape. In
Figure~\ref{fig3} we depict $\vec{d}(k)$ for fixed values of $\mu=1$,
$\Delta=1$ and three different values $t=0.25,0.5,1$ with the
corresponding energy dispersion relation $E(k)$.  If the origin of the
$(y,z)$ plane is included in the closed contour that the ellipse
describes, its winding number is one, $W=1$. If not, the winding
number is zero, $W=0$. These are two topologically separated
scenarios, reflecting the distinctly different role of the Majorana
fermions in the top and bottom parts of Figure~\ref{fig2}. For $W=0$,
the superconducting phase is the normal or trivial one, while it is
topological for $W=1$. A special situation occurs if the ellipse just
touches the $x$ axis, the band gap disappears, marking the phase
transition point.

\subsection{Chirality, Random Winding Numbers and Modeling Aspects}
\label{sec23}

When studying such topological invariants in statistical physics, the
closed contour might be a random quantity, for example, generated by a
proper ensemble of Hamiltonians. In the case of the Kitaev chain, this
ensemble may be realized by choosing the parameters $\mu$, $\Delta$
and $t$ from probability distributions. Hence, the contour can be
different for a particular choice, i.e.~it becomes random, and the
winding number $W$ will be random as well. In general, the dynamics of
a system under consideration, described by the Hamiltonian, and the
distributions of its parameters will determine the distribution
$P(W)$. Are there universalities when comparing different systems? ---
If yes, in which quantities do these universalities manifest? --- In
the distributions $P(W)$ on their original scales, or on some scales
which make these systems comparable? --- These are the guiding
questions for our research. Universalities are best identified in
random schematic systems that only contain the most basic ingredients
needed for the relevant physics, in the present case for the occurence
of winding numbers.  Random Matrix Theory (RMT)
\cite{MehtaBook,GuhrRMTReview} is known to be a powerful concept in
this spirit when studying universalities in spectral correlations as
well as in the correlations of parametric level motion
\cite{SimonsAltshuler1993A,SimonsAltshuler1993B}.  The chiral symmetry
\eqref{eq1.6} and thus the restriction to two dimensions are essential
for the interpretation of the two superconducting phases in the Kitaev
chain in terms of the winding number. Hence, when setting up a
schematic random matrix model, we need to employ chirality.

\subsection{Connections to Quantum Chromodynamics and Condensed Matter Physics}
\label{sec24}

In Quantum Chromodynamics the chiral symmetry of the Dirac
operator is broken spontaneously as well as explicitly by the quark
masses. The chiral condensate is the order parameter of the phase
transition that occurs at a high temperature and that restores chiral
symmetry, which is related to the confinement--deconfinement
transition. To investigate statistical properties of lattice gauge
calculations, chiral RMT
\cite{Verbaarschot1994,VerbaarschotWettig2000,ShuryakVerbaarschot1993,
  Wettig1996A, Wettig1996B, JacksonVerbaarschot1996,
  VerbaarschotZahed1993,GWW2000} is remarkably successful.  As in
original RMT, presence or absence of time--reversal invariance
combined with spin-rotation symmetries results in three classes of
chiral random matrices: orthogonal, unitary, and symplectic. It was
then shown that altogether ten RMT symmetry classes
\cite{AltlandZirnbauer1997, Heinzner2005, Kitaev2009, Schnyder2008, Ryu2016} exist,
referred to as the tenfold way. The three original and the three
chiral ones comprise six of these ten classes, the remaining four
emerge when particle--hole symmetry is also considered, see
Refs.~\cite{Oppermann1990,Zirnbauer2021}.  In condensed matter physics
chiral symmetry is realized by sublattice symmetry (see early work in
Ref.~\cite{Gade1993}) or as a combination of time reversal and
particle-hole symmetry \cite{Zirnbauer2021}.

In the terminology of condensed matter physics, the winding number
comes in as characterization of translationally invariant
one--dimensional chiral systems that are gapped at the center of the
spectrum. The winding number is the integer topological index with
respect to the bundle of negative--energy bands. A non--zero winding
number $W$ indicates the topologically nontrivial situation with $|W|$
modes at each boundary
\cite{prodanBulkBoundaryInvariants2016,ChenChiou2020,
  Shapiro2020,Alicea2012}. The winding number differs for different
realization of the disorder, i.e.~it becomes random. Our research on
the winding number was inspired by studies of systems with energy
bands in two dimensions, allowing for a topological classification by
the (first) Chern number. A random matrix model
\cite{WalkerWilkinson1995,GatWilkinson2021} revealed a Gaussian
distribution of Chern numbers with a universal covariance.

\section{Formulation of the Problem and Mathematical Setup}
\label{sec3}

After introducing chiral random matrix ensembles with a parameter
dependence in Section~\ref{sec31}, the statistical quantities of
interest are defined in Section~\ref{sec32}. In Section~\ref{sec33} a
crucial step for all of our mathematical investigations is presented,
namely the mapping of the topological problem addressed to a spectral
one which greatly facilitates the computations.

\subsection{Chiral Random Matrix Ensembles with Parametric Dependence}
\label{sec31}

We derived results \cite{Braun2022,Hahn2022} for the chiral unitary
and the chiral symplectic symmetry classes labeled AIII and CII,
respectively, see Ref.~\cite{AltlandZirnbauer1997}.  The latter case
is mathematically much more demanding than the former, but not as
involved as the orthogonal case, labeled BDI. The cases BDI and CII
describe time--reversal invariant systems, while this invariance does
not exist in the case AIII. We refer to the matrices as Hamiltonians
$H$, as most of the present application of winding numbers seem to
stem from Condensed Matter Physics.  The matrices $H$ are complex
Hermitean or quaternion real, i.e.~self--adjoint, with even dimension
$\beta N\times \beta N$ where we employ the Dyson indices $\beta=2$
and $\beta=4$ for AIII and CII.  Chiral symmetry manifests in the
relation
\begin{equation} \label{2ChiralSymmetry}
\{\mathcal{C},H\} = 0
\end{equation}
where in the chiral basis
\begin{equation}
\mathcal{C} = \begin{bmatrix}
\eins_{\beta N/2} & 0
\\
0 & -\eins_{\beta N/2} 
\end{bmatrix} \ .
\end{equation}
The Hamiltonians thus take the block off--diagonal form
\begin{equation} \label{2ChiralHamiltonian}
H = \begin{bmatrix}
0 & K
\\
K^\dag & 0
\end{bmatrix} \ ,
\end{equation}
where the $\beta N/2 \times \beta N/2$ matrices $K$ have no further
symmetries.  We draw the matrices $H$ form the chiral Gaussian
Unitary, respectively, symplectic Ensembles (chGUE, chGSE). To study
questions of topology, we give these random matrices a parametric
dependence $K=K(p)$ and thus $H=H(p)$, where the real variable $p$
lies on the unit circle. The winding number corresponding to these
Hamiltonians is then \cite{Maffei2018, AsbothBook}
\begin{equation} \label{2WindingNumberDef}
W = \frac{1}{2\pi i} \int\limits_0^{2\pi} w(p) \, dp \ ,
\end{equation}
with the winding number density
\begin{equation} \label{2WindingNumberDensityDef}
w(p) = \frac{d}{dp} \ln \det K(p) = \frac{1}{\det K(p)} \frac{d}{dp} \det K(p) \ .
\end{equation}
Cauchy's argument principle applies to the integral
\eqref{2WindingNumberDef}, provided $\det K$ is a nonzero analytic
function of $p$, see Section~\ref{sec21} and particularly
Eq.~\eqref{eq1.3}.

To produce explicit results, we choose a particular realization of the
parameter dependence.  With two smooth and $2\pi$ periodic scalar
functions $a(p)$ and $b(p)$, we set
\begin{equation} \label{2RMF}
K(p) = a(p) K_1 + b(p) K_2 \ ,
\end{equation}
where the matrices $K_1$ and $K_2$ have dimensions $\beta N/2\times
\beta N/2$. The associated Hamiltonians 
\begin{equation} \label{2RMFH}
H(p) = a(p) H_1 + b(p) H_2  \qquad \text{with} \qquad H_m = \begin{bmatrix}
0 & K_m
\\
K_m^\dag & 0
\end{bmatrix} \ , \  m=1,2 \ , 
\end{equation}
define parametric combinations of either two chGUE's or two chGSE's.
Averages over these combined ensembles have to be performed. It is
convenient to introduce the vector
\begin{equation}\label{2vdef}
v(p) = (a(p),b(p)) \ \in\mathbb{C}^2 \ .
\end{equation}
Time--reversal invariance imposes the condition $v^*(p) = v(-p)$ in
the chiral symplectic case CII.

\subsection{Statistical Quantities Considered}
\label{sec32}

Considering $k$ different points $p_i, \ i=1,\ldots,k$, on the unit
circle, we are interested in the $k$--point correlators of winding
number densities
\begin{equation} \label{2kPointCorrelationDef}
C^{(\beta,N)}_k (p_1,\ldots,p_k) = \left\langle w(p_1) \cdots w(p_k) \right\rangle
\end{equation}
The precise meaning of the angular brackets indicating the ensemble
average will be given later on.  In the chiral unitary case AIII, we
computed these correlators directly \cite{Braun2022}, see
Section~\ref{sec41}. As, first, this approach becomes forbiddingly
complicated in the chiral symplectic case CII, and, second, results in
cumbersome expressions for larger $k$, we calculated the generators
\begin{equation} \label{genfct}
Z^{(\beta,N)}_{k|l}(q,p) = \left\langle \frac{\prod_{j=1}^l \det K(p_j)}{\prod_{j=1}^k \det K(q_j)} \right\rangle
\end{equation}
for two sets of variables $p_1,\ldots,p_l$ and $q_1,\ldots,q_k$ in
Ref.~\cite{Hahn2022}, see Section~\ref{sec44}. Only the case $k=l$ is
needed, but the more general definition~\eqref{genfct} for $k$ and $l$
being different has technical advantages.  We notice that $k$ and $l$
are the numbers of determinants in denominator and numerator,
respectively.  The $k$--fold derivative
\begin{equation} \label{2kPointCorrelationGen}
C^{(\beta,N)}_k(p_1,\ldots,p_k) = \frac{\partial^k}{\prod_{j=1}^k \partial p_j} Z^{(\beta,N)}_{k|k}(q,p)\Bigg|_{q=p}
\end{equation}
of the generator~\eqref{genfct} for $k=l$ at $q=p$ yields the
correlator \eqref{2kPointCorrelationDef}. Anticipating the later
discussion, we emphasize that the generators for both Dyson indices
$\beta=2,4$ will exhibit a remarkably clear structure \cite{Hahn2022}
which is an important reason to address them here. It is worth
mentioning that the correlators \eqref{2kPointCorrelationDef} and the
generators \eqref{genfct} are very different from those for the
parametric level motion considered in
Refs.~\cite{SimonsAltshuler1993A,SimonsAltshuler1993B}.

Furthermore, we also computed the distribution of winding numbers
$P(W)$ in the chiral unitary case AIII \cite{Braun2022}, see
Section~\ref{sec42}.

\subsection{Mapping a Topological to a Spectral Problem}
\label{sec33}

At first sight, the computation of the correlators
\eqref{2kPointCorrelationDef} and the generators \eqref{genfct}
appears as a formidable task, requiring the development of completely
new techniques in RMT. Luckily, one can establish a link between the
topological problem set up above and spectral problems in RMT for
which a wealth of literature exists. This amounts to a tremendous
simplification, even though the calculations to be carried out are
still involved and quite demanding particularly in the chiral
symplectic case.  The key observation is that a combination
of the two matrices $K_1$ and $K_2$ in Eq.~\eqref{2RMF} encodes all the
statistical information needed. Pulling out $K_1$, say, one has
\begin{equation} \label{repar}
K(p)=a(p)K_1 + b(p)K_2 = b(p)K_1\left(\kappa(p)\eins_{\beta N/2}+K_1^{-1}K_2\right) 
\end{equation}
with the ratio
\begin{equation} \label{reparkappa}
\kappa(p)=\frac{a(p)}{b(p)} \ .
\end{equation}
Since the winding number density \eqref{2WindingNumberDensityDef} is the derivative
of the logarithm
\begin{equation} \label{pulldet}
\ln \det K(p) = \ln \det K_1 + \beta N \ln b(p) + \ln \det \left( \kappa(p) + K_1^{-1}K_2 \right) \ ,
\end{equation}
the first term $\ln \det K_1$ does not contribute and, remarkably,
only the combination $Y=K_1^{-1}K_2$ is relevant. Using Eq.~\eqref{repar} the
generators acquire the form
\begin{equation} \label{genfctrepar}
  Z^{(\beta,N)}_{k|k}(q,p) =  \left( \prod_{j=1}^k \frac{b(p_j)}{b(q_j)} \right)^{\beta N}
  \left\langle \prod_{j=1}^k \frac{\det(\kappa(p_j)\eins_{\beta N/2}+Y)}{\det(\kappa(q_j)\eins_{\beta N/2}+Y)} \right\rangle \ ,
\end{equation}
which as well only contains the matrix $Y$.

The task to be solved is the derivation of the probability density for
the random matrices $Y=K_1^{-1}K_2$ from the independent Gaussian
distributions for the random matrices $K_1$ and $K_2$.  Once again luckily,
the results are known as spherical~\cite{Krishnapur2009,Mays2013}
ensembles and their probability densities read explicitly
\begin{equation} \label{spherical.beta}
\widetilde{G}^{(\beta)}(Y) = \frac{1}{\pi^{\beta N^2/2}} \prod_{j=1}^N \frac{\left(\beta(N+j)/2-1\right)!}{\left(\beta j/2-1 \right)!}\
            \frac{1}{\det^{2 N}\left( \eins_{\beta N/2} + YY^\dagger \right)} \ .
\end{equation}
These ensembles are referred to as complex spherical and quaternion
spherical for $\beta=2,4$.  In the complex case, the probability
density \eqref{spherical.beta} can be reduced to a joint probability
density of the $N$ complex eigenvalues $z={\rm
  diag\,}(z_1,\ldots,z_{N})$ of $Y$ and reads
\begin{equation} \label{sphericalev.beta2}
  G^{(2)}(z)  = \frac{1}{c^{(2)}} |\Delta_N(z)|^2 \prod_{j=1}^N \frac{1}{(1+|z_j|^2)^{N+1}}
\end{equation}
with the the Vandermonde determinant
\begin{equation} \label{Vandermonde}
  \Delta_N(z) = \prod_{j<l} (z_j-z_l) \ .
\end{equation}
In the quaternion case, however, each eigenvalue $z_j$ of $Y$ has a
complex conjugate $z_j^*$ which is also an eigenvalue. The
corresponding joint probability density of the eigenvalues $z={\rm
  diag\,}(z_1,z_1^*,z_2,z_2^*,\ldots,z_N,z_N^*)$ is given by
\begin{equation} \label{sphericalev.beta4}
  G^{(4)}(z)  = \frac{1}{c^{(4)}} \Delta_{2N}(z) \prod_{j=1}^N \frac{z_j-z_j^*}{(1+|z_j|^2)^{2N+2}} \ .
\end{equation}
The normalization constants are
\begin{equation} \label{normbeta}
c^{(\beta)} = (\beta\pi/2)^NN!\prod_{j=1}^N{\rm B}(\beta j/2,\beta(N+1-j)/2) \ ,
\end{equation}
where $\textrm{B}(x,y)$ is Euler's Beta function.  The question
whether the integrals to be done are well--defined for $\beta=4$
arises, but the answer is affirmative \cite{Hahn2022}. Hence, the
ensemble average over a function $f(z)$ to be performed amounts to
carrying out the integral
\begin{equation} \label{zaverage}
\langle f(z) \rangle = \int\limits_\mathbb{C} d[z_1] \cdots  \int\limits_\mathbb{C} d[z_N] \,  G^{(\beta)}(z) \, f(z) 
\end{equation}
over all complex eigenvalues. Hence, by reducing the two chiral
ensembles to a single spherical one for either $\beta$, all
information of the topological problem is contained in the
determinants $\det(\kappa(p)\eins_{\beta N/2}+Y)$ or their
derivatives. Most advantageously, this is equivalent to a spectral
problem where $Y$ and $\kappa(p)$ formally play the roles of a
(complex or quaternion) ``Hamiltonian'' and of the corresponding ``energy'',
respectively.

\section{Results}
\label{sec4}

The correlators for the unitary case are addressed in
Section~\ref{sec41}, the distribution is given in in
Section~\ref{sec42}. Aspects of universality are discussed in
Section~\ref{sec43}.  The generators in the chiral unitary and
symplectic cases are presented in Section~\ref{sec44}.

\subsection{Winding Number Correlators in the Chiral Unitary Case}
\label{sec41}

In Ref.~\cite{Braun2022}, we calculated the winding number correlators 
$C^{(2,N)}_k (p_1,\ldots,p_k)$ as defined in Eq.~\eqref{2kPointCorrelationDef}
in the unitary case directly. We chose
\begin{equation} \label{abchoice}
a(p) = \cos p \qquad \text{and} \qquad b(p) = \sin p \ .
\end{equation}
Using Eqs.~\eqref{2WindingNumberDensityDef} and \eqref{pulldet} as
well as the complex eigenvalues of $Y$, one has
\begin{equation} \label{3WindingNumberDensity}
w(p) = N\cot p + y(p)
\qquad \text{with} \qquad
y(p) = - \frac{1}{\sin^2 p}\sum_{n=1}^N \frac{1}{\cot p + z_n} \ .
\end{equation}
Only the $k$--fold products of $y(p)$ have to be ensemble averaged
with the joint probability density \eqref{sphericalev.beta2}, the
presence of the inconvenient term $N\cot p$ implies that the correlator
$C^{(2,N)}_k (p_1,\ldots,p_k)$ of the $k$  winding number densities
$w(p_j)$ becomes a combinatorial sum of the $y(p_j)$ correlators.
Furthermore, the latter themselves turn out to be rather involved
combinatorial expressions. Eventually, $C^{(2,N)}_k (p_1,\ldots,p_k)$
is found to be a combinatorial sum of determinants with the entries
\begin{equation}\label{FunctionL}
L_{nml}(q_l) = \frac{(-1)^{m-n} \pi}{q_l^{m-n+1}} \textrm{B}(m,N-m+1)
\begin{cases}
u_m(N,q_l^2) \qquad &m\geq n
\\
-v_m(N,q_l^2) \qquad &m<n
\end{cases} \ ,
\end{equation}
with the properly normalized incomplete Beta functions
\begin{eqnarray} \label{3ReducedBetaFunctions}
u_m(N,q_l^2) &=& \frac{2}{\textrm{B}(m,N-m+1)} \int\limits_0^{q_l} d\rho \frac{\rho^{2m-1}}{(1+\rho^2)^{N+1}} \nonumber\\
v_m(N,q_l^2) &=& \frac{2}{\textrm{B}(m,N-m+1)} \int\limits_{q_l}^\infty d\rho \frac{\rho^{2m-1}}{(1+\rho^2)^{N+1}}
\end{eqnarray}
that satisfy $u_m(N,q_l^2) + v_m(N,q_l^2) = 1$. Even though
$\textrm{B}(m,N-m+1)$ drops out in the $L_{nml}(q_l)$, this
normalization has advantages, see Ref.~\cite{Braun2022}. The first two
correlators read
\begin{eqnarray} \label{3kPointCorrelationResults}
C^{(2,N)}_1 (p_1) &=& 0  \nonumber\\
C^{(2,N)}_2 (p_1,p_2) &=& -\frac{1-\cos^{2N} \left(p_1 - p_2\right)}{1-\cos^2 \left(p_1 - p_2\right)} \ .
\end{eqnarray}
The at first sight surprising vanishing of the averaged winding number density is actually quite
natural, as the winding number $W$ must have a symmetric distribution with vanishing first
moment. The integral of $C^{(2,N)}_1 (p_1)$ over $p_1$ is this first moment.

\subsection{Winding Number Distribution}
\label{sec42}

In Ref.~\cite{Braun2022}, we also computed the winding number
distribution $P(W)$ in the unitary case for the choice
\eqref{abchoice}. Using Cauchy's argument principle, we derive the
discrete probability distribution
\begin{equation} \label{3WindingNumberDistribution}
P(W) = r\left(\frac{W+N}{2}\right) \binom{N}{(W+N)/2} 
\end{equation}
on the integers $W$ between $-N$ and $N$ for arbitrary, finite matrix
dimension $N$. Here, $r(m)$ is the probability that $m$ eigenvalues
are inside the unit circle and the remaining ones outside which may be
written as
\begin{equation} \label{3qProbabilityD}
r(m) = \int\limits_{|z_1| < 1} d[z_1] \cdots \int\limits_{|z_m| < 1} d[z_m]
        \int\limits_{|z_{m+1}| > 1} d[z_{m+1}] \cdots \int\limits_{|z_N| > 1} d[z_N] \, G^{(2)}(z) \ .
\end{equation}
Doing the integrals yields
\begin{equation} \label{3qProbabilityR}
r(m) = \frac{1}{N!} \sum_{\omega \in \mathbb{S}_N} \left(\prod_{i=1}^m u_{\omega(i)}(N,1)\right)
                 \left(\prod_{i=m+1}^N v_{\omega(i)}(N,1)\right),  
\end{equation}
in terms if the functions~\eqref{3ReducedBetaFunctions}. The
combinatorial factor in formula~\eqref{3WindingNumberDistribution}
takes into account the permutation invariance of the eigenvalues
inside, respectively outside, the unit circle. The sum runs over all
permutations, $\mathbb{S}_N$ is the permutation group.

\subsection{Aspects of Universality}
\label{sec43}

The quest for universality is twofold, first, there is the question
whether the same statistical effects, distributions or scalings, etc,
can be identified in empirical or experimental data of different
physical systems. Second, there is the theoretical and mathematical
side concerned with often schematic models and their ability to
describe or even predict the results from data analysis. In the case
of spectral correlations, universal statistics is found on the local
scale of the mean level spacing, i.e.~universalities are revealed
after a rescaling of the energies, referred to as unfolding. The
unfolded correlators of, on the one hand, RMT for infinite level
number and of, on the other hand, numerous physical systems of very
different nature with large number of levels coincide, see the
discussion in Refs.~\cite{GuhrRMTReview,MehtaBook}. The theoretical
and mathematical challenge is non--trivial as it amounts to showing
that a most general class of probability densities for the random
matrices yields after unfolding the same statistical quantities. Put
differently, it suffices to consider Gaussians, because the resulting
statistics is, always after unfolding, universal.

In the case of statistical topology, universality is of equally high
importance, but appears to be considerably more complicated. Already
on the theoretical and mathematical side there are several natural
questions to be posed: First, is there an unfolding scale comparable
to the local mean level spacing and how is it related to the scale of
the level velocity as in the parametric correlations
\cite{SimonsAltshuler1993A, SimonsAltshuler1993B,BeenakkerRejaei1994}?
--- Second, which probability densities for the random matrices yield
in the model set up in Section \ref{sec31} the same statistics? ---
Third, what are the conditions on the functions $a(p)$ and $b(p)$ or,
more precisely, the combined conditions on these functions and the
probability densities that yield in the model universal statistics?
--- Fourth, is it possible to find universal statistics for models
more general than the one of Section \ref{sec31}?

In Ref.~\cite{Braun2022}, we started addressing these issues in the
unitary case for the choice \eqref{abchoice}. Guided by
unfolding in spectral statistics, we rescaled the arguments $p_i$
in the correlation functions $C^{(2,N)}_k (p_1,\ldots,p_k)$
according to
\begin{equation} \label{3Rescaling}
\psi_i = N^\alpha p_i \ . 
\end{equation}
The power $\alpha$ should be positive, because we want to zoom into
\begin{figure}[htbp]
\centering
\includegraphics[width=.9\linewidth]{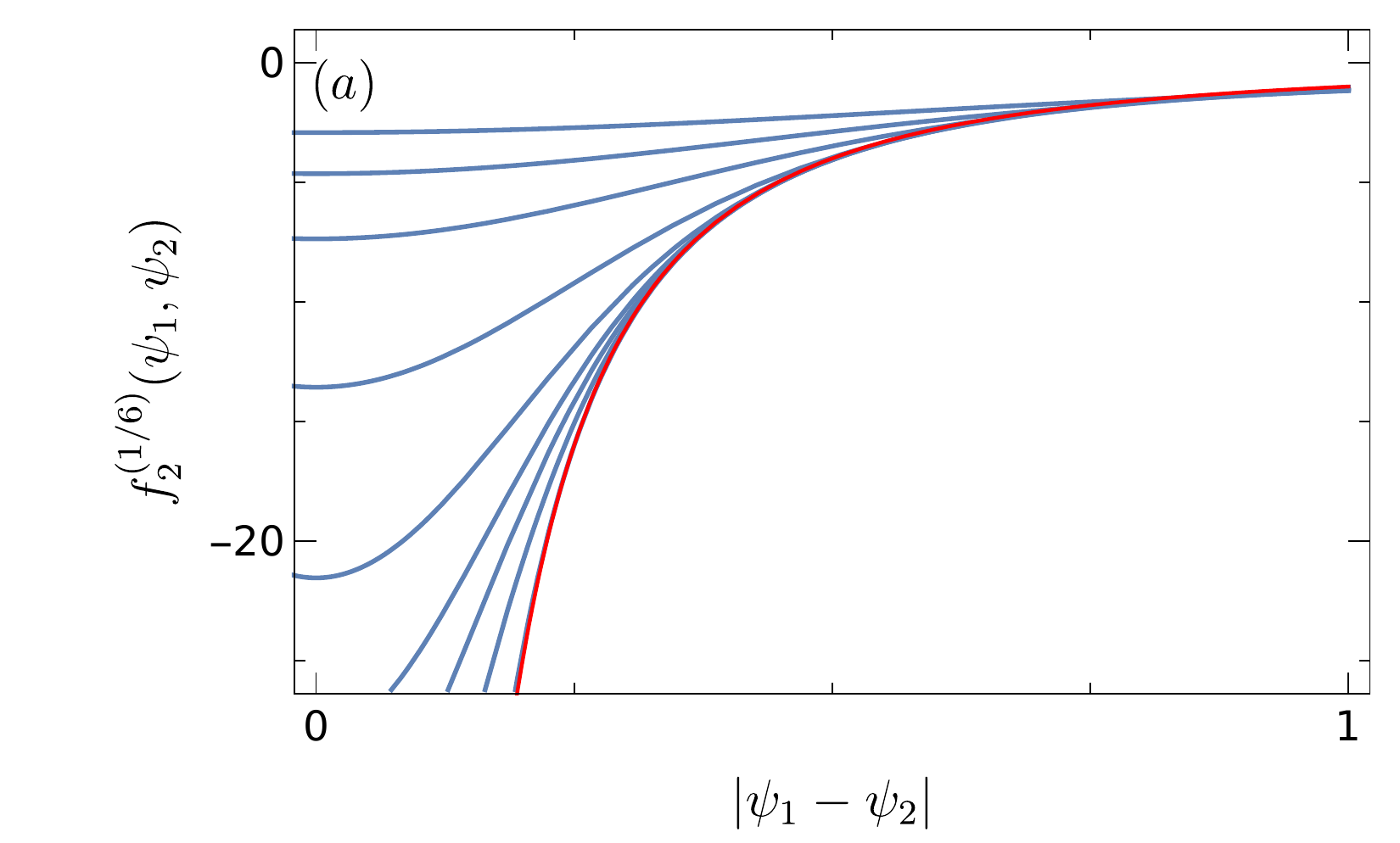}
\\
\includegraphics[width=.9\linewidth]{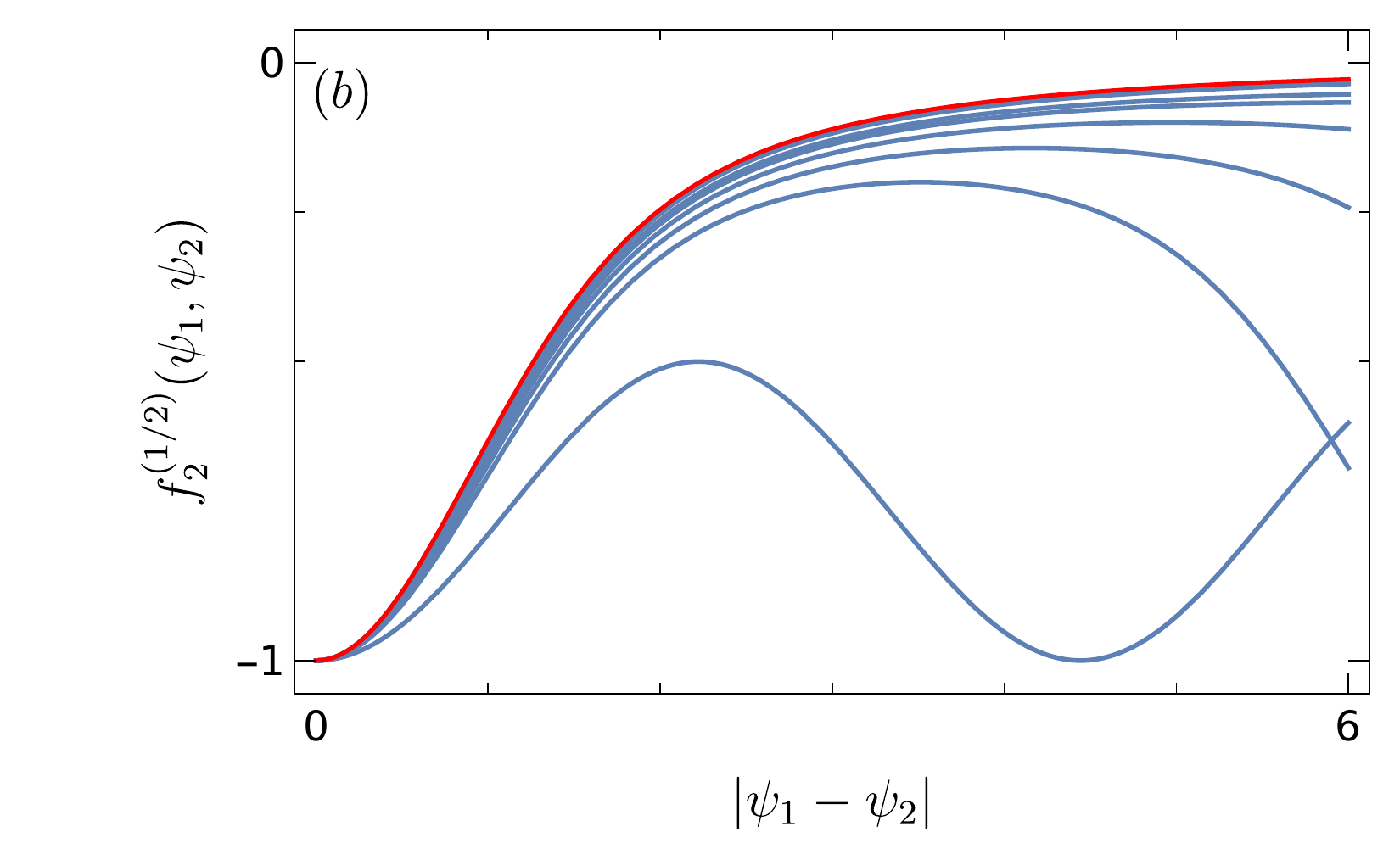}
\caption{Unfolded two-point function after the rescaling
  \eqref{3Rescaling} for different values of $N$ (blue). In (a) we
  used $N = 5, 10, 20, 50, 100, 150, 200, 300, 1000$ and $\alpha =
  1/6$, in (b) $N = 2, 5, 7, 10, 15, 20, 50, 100$ and $\alpha =
  1/2$. For comparison the limit \eqref{3UnfoldingLimit} (red). Taken
  from Ref.~\cite{Braun2022}}
\label{fig1Braun2022}
\end{figure}
the parametric dependence in the limit $N\rightarrow\infty$. Consider
the two--point function \eqref{3kPointCorrelationResults} and the
limit
\begin{equation} \label{3UnfoldingLimit}
  \lim_{N\rightarrow \infty} C^{(2,N)}_2\left(\frac{\psi_1}{N^\alpha},\frac{\psi_2}{N^\alpha}\right)
             \frac{d\psi_1}{N^\alpha} \frac{d\psi_2}{N^\alpha} = f_2^{(\alpha)}(\psi_1,\psi_2) d\psi_1 d\psi_2
\end{equation}
defining the function $f_2^{(\alpha)}$, if existing.  A
straightforward calculation yields
\begin{equation}\label{3UnfoldingLimitB}
f_2^{(\alpha)}(\psi_1,\psi_2) = \begin{cases}
\displaystyle -\frac{1}{\left(\psi_1 - \psi_2\right)^2} \qquad &\alpha < \frac{1}{2}
\\
\displaystyle -\frac{1-\exp(-( \psi_1 - \psi_2)^2)}{\left( \psi_1 - \psi_2\right)^2} \qquad &\alpha = \frac{1}{2}
\\
\displaystyle 0 \qquad &\alpha > \frac{1}{2}
\end{cases} \ .
\end{equation}
We notice $C^{(2,N)}_2(p_1,p_1)=-1$, see
Eq.~\eqref{3kPointCorrelationResults}, implying that
$\psi_1\neq\psi_2$ when taking the limit for arbitrary $\alpha$.  The
result \eqref{3UnfoldingLimitB} reveals different regimes, the one for
$\alpha = 1/2$ involves the same scale as in
Refs.~\cite{SimonsAltshuler1993A, SimonsAltshuler1993B}.  
Figure~\ref{fig1Braun2022} shows results for two values of $\alpha$
and various values of $N$, the unfolded two-point function
approaches the limit \eqref{3UnfoldingLimitB} when $N$ increases. We
conjectured that the function $f_2^{(\alpha)}(\psi_1,\psi_2)$ is
universal~\cite{Braun2022}.

In Ref.~\cite{Braun2022}, we also showed that the winding number
distribution \eqref{3WindingNumberDistribution} becomes Gaussian for
large $N$. More precisely, its second moment behaves like $\langle
W^2\rangle \sim \sqrt{N}$, suggesting an unfolding of the form
$W/N^{1/4}$, i.e.~different from the rescaling above. It then follows
that $P(W)$ approaches a Gaussian with variance $2\sqrt{N/\pi}$ for
large $N$.

\subsection{Generators in the Chiral Unitary and Symplectic Cases}
\label{sec44}

We computed the generators \eqref{genfct}, respectively
\eqref{genfctrepar} exactly for $\beta=2$ and $\beta=4$ in
Ref.~\cite{Hahn2022}. To this end, we used the method put forward some
years ago in Refs.~\cite{KieburgGuhr2010a,KieburgGuhr2010b}. It
identifies and employs, in ordinary space, supersymmetric structures
deeply rooted in the ensemble averages. As there is no mapping
performed of the ensemble averages to superspace, the method is often
referred to, jokingly, but not deceptively, as ``supersymmetry without
supersymmetry''. In the chiral unitary case $\beta=2$, we found a
ratio of two determinants,
\begin{equation} \label{genfctAIII}
  Z^{(2,N)}_{k|k}(q,p) = \frac{\det \left[\displaystyle
      \frac{1}{v^T(q_m)\sigma_2v(p_n)}\left(\frac{v^\dagger(q_m) v(p_n)}{v^\dagger(q_m) v(q_m)}\right)^N \right]_{1\leq m,n\leq k}}
            {\det \left[\displaystyle \frac{1}{v^T(q_m)\sigma_2v(p_n)} \right]_{1\leq m,n\leq k}} \ ,
\end{equation}
where $\sigma_2$ is the second $2\times 2$ Pauli matrix and $v(p_n)$
te vector defined in Eq.~\eqref{2vdef}.  In the chiral symplectic case
$\beta=4$, we arrived at a ratio of a Pfaffian and a determinant,
\begin{equation} \label{genfctCII}
Z^{(4,N)}_{k|k}(q,p) = \frac{\Pf \begin{bmatrix}\displaystyle
\widehat{\rm K}_1(p_m,p_n) & \widehat{\rm K}_2(p_m,q_n) \\
-\widehat{\rm K}_2(p_n,q_m) &  \widehat{\rm K}_3(q_m,q_n) \end{bmatrix}_{1\leq m,n\leq k}}
{\det \left[\displaystyle \frac{1}{iv^T(q_m)\sigma_2v(p_n)} \right]_{1\leq m,n\leq k}} \ .
\end{equation}
The three kernel functions $\widehat{\rm K}_l(p_m,p_n) \ , l=1,2,3$
are quite complicated and can be found explicitly in
Ref.~\cite{Hahn2022}. Considering the complexity of the problem and of
its mathematical structure, these are remarkably compact results, even
in the chiral symplectic case. This compactness is the reason why we
give these results here. Their form is intimately connected with the
mapping of the topological to a spectral problem discussed in
Section~\ref{sec33}, because such determinant and Pfaffian expressions
are ubiquitous for the generators in spectral statistics. Importantly
this carries over, at least for the model considered, to the
generators for the correlators of winding number densities.

\section{Discussion and Conclusions}
\label{sec5}

Statistical Topology is an emerging branch in statistical physics,
with connections to various branches of mathematics. It is triggered
by the identification of topological questions in many areas of
physics, ranging from quantum mechanics and quantum field theory over
semiclassics to QCD and Condendsed Matter Physics. First, I tried to
give an introduction to winding number statistics for newcomers who do
not have any background, avoiding usage of expert jargon and of
burying the key ideas under the adavanced terminology developed in
mathematics and mathematical physics. Second, I reviewed results that
my collaborators and I obtained in two recent works. We studied
winding numbers and associated statistical quantities in a random
matrix model.  There are, of course, also other topological invariants
of considerable interest in physics, most notably the Chern numbers.

I presented our first, probably awkward, steps to look at univeral
behavior. In my opinion, the most fascinating challenge for the future
is the further study of universality in statistical topology, more
precisely, of both of its aspects, the experimental--empirical as well
as the theoretical--mathematical one.

\section*{Acknowledgements}

I thank Omri Gat, Nico Hahn, Mario Kieburg and Daniel Waltner, my
collaborators of Refs.~\cite{Braun2022,Hahn2022}, I deeply regret that
I cannot thank anymore Petr Braun who passed away in late 2020. I am
greatful to Nico Hahn for Figure~\ref{fig3}.  This work was funded by
the German--Israeli Foundation within the project \textit{Statistical
  Topology of Complex Quantum Systems}, grant number GIF
I-1499-303.7/2019.

\bibliographystyle{mystyle}

\bibliography{strc_extended}

\end{document}